\documentclass[10pt]{revtex4}
\usepackage{amssymb,epsf}
\usepackage{latexsym}
\usepackage{graphicx}
\usepackage{color}

\begin{document}

\title{Phantom   Wormhole Solutions\\ in a Generic Cosmological Constant Background}
\author{Y. Heydarzade$^1$\footnote{heydarzade@azaruniv.edu}, N. Riazi$^2$\footnote{n-riazi@sbu.ac.ir}
and H. Moradpour$^3$\footnote{h.moradpour@riaam.ac.ir}}
\affiliation{$^1$ Department of Physics, Azarbaijan Shahid Madani
University, Tabriz, 53714-161 Iran\\$^2$ Department of Physics,
Shahid Beheshti University,
Evin, Tehran 19839, Iran\\
$^3$ Research Institute for Astronomy and Astrophysics of Maragha
(RIAAM), P.O. Box 55134-441, Maragha, Iran.}

\begin{abstract}
There are a number of reasons to study wormholes with generic
cosmological constant $\Lambda$. Recent observations indicate that
present accelerating expansion of the universe demands
$\Lambda>0$. On the other hand, some extended theories of
gravitation such as supergravity and superstring theories posses
vacuum states with $\Lambda<0$. Even within the framework of
general relativity, a negative cosmological constant permits black
holes with horizons topologically different from the usual
spherical ones. These solutions are convertible to wormhole
solutions by adding some exotic matter.  In this paper, the asymptotically
flat wormhole solutions in a generic
cosmological constant background are studied. By constructing a specific
class of shape functions, mass function, energy density and
pressure profiles which support such a geometry are obtained. It
is shown that for having such a geometry, the wormhole throat
$r_0$, the cosmological constant $\Lambda$ and the equation of
state parameter $\omega$ should satisfy two specific conditions.
The possibility of setting different values for the parameters of
the model helps us to find exact solutions for the metric
functions, mass functions and energy-momentum profiles. At last,
the volume integral quantifier, which provides useful information
about the total amount of energy condition violating matter is
discussed briefly.

\end{abstract}

\maketitle

\section{Introduction}
In general relativity, geometrical bridges connecting two distant
regions of a universe or even two different universes are in
principle possible. Spacetimes containing such bridges appear  as
solutions of the Einstein field equations. The term '"wormhole"
for these bridges was used for the first time in 1957 by J. A.
Wheeler \cite{Wheeler1, Wheeler2}. Many years later in 1988,  the
notion of traversable Lorentzian wormholes attracted the attention
of  physicists by the fundamental papers of Morris, Thorne and
Yurtsewer \cite{Thorne1, Thorne2}. In these papers it was shown
that such wormholes could allow humans not only to travel between
universes, or distant parts of the same universe, but also to
construct time machines. Also, It has been suggested that black
holes and wormholes are interconvertible structures and stationary
wormholes could be possible as final states of black-hole
evaporation \cite{Hayward}. Moreover, it is shown that
astrophysical accretion of ordinary matter could convert wormholes
into black holes \cite{ Kardashev, Kuhfittig, Sushkov-Zaslavskii}.
In the wormhole physics, it is known that these structures do not
satisfy common energy conditions. The energy-momentum tensor  of
the matter supporting such  geometries violates the null energy
condition at least in the vicinity of the wormhole throat
\cite{Hochberg1, Hochberg2, Hochberg3}. The matter that violates
the null energy condition is usually called as exotic matter.
Since the violation of the energy conditions is conventionally
considered as a problematic issue, minimizing its usage seems to
be useful. One may obtain that in the context of thin-shell
wormholes using the cut-and-paste procedure \cite{Visser1,
Garcia}. In this context, the exotic matter is concentrated at the
throat of the wormhole, which is localized on the thin shell.

Another approach lies within modified theories of gravity, where
normal matter threading the wormhole satisfies the energy
conditions, and they are the higher order curvature terms that
support these exotic geometries. In the context of modified
gravity, the gravitational field equation may be written as
$G_{\mu\nu} \equiv\ R_{\mu\nu}-\frac{1}{2}Rg_{\mu\nu}=8\pi G
T_{\mu\nu}^{eff}$, where $T_{\mu\nu}^{eff}$ is the effective
energy-momentum tensor. Then, in modified theories of gravity, the
effective energy-momentum tensor involving higher order
derivatives violates the null energy condition, i.e.,
$T_{\mu\nu}^{eff}k^{\mu}k^{\nu}<0$ where $k^{\mu}$ is a null
vector. This approach is widely analysed in the literature as in
the fame work of  $f(R)$ gravity \cite{Oliveira, Sajadi},
curvature matter couplings \cite{Garcia2, Garcia3}, conformal Weyl
gravity \cite{Lobo1} and braneworlds \cite{Lobo2}. On the other
hand, according to the recent discoveries in  cosmology, our
universe is in  accelerated expansion \cite{a1,a2,a3,a4}. A
dominating dark energy component with an equation of state
$p=\omega \rho$ with $\omega<-\frac{1}{3}$, is thought to be
responsible for this accelerated expansion phase of  universe. The
specific ranges of $\omega<-1$, $\omega=-1$ and $-1<\omega<-1/3$
correspond to the phantom energy, cosmological constant and
quintessence matter, respectively. Then, one of the reasons to
study wormholes with generic cosmological constant $\Lambda$ and
specially $\Lambda>0$ turns to the accelerated expansion of the
universe.  Another reason to investigate wormhole solutions with
generic cosmological constant $\Lambda$ turns to supergravity and
superstring theories which have vacuum states with $\Lambda<0$.
Also, in the framework of general relativity, a negative
cosmological constant allows black hole solutions with horizons
that are topologically different from the usual spherical ones.
Adding some exotic matter can convert these black hole solutions
to wormhole solutions \cite{Lambda1, Lambda2}.

In addition, the phantom energy possess some special features such
as a divergent cosmic scale factor in a finite time
\cite{caldwell1, caldwell2}, leading to appearance of negative
entropy and temperature \cite{Brevik, Nojiri} and predicting a new
long range force \cite{Amendola}. Since the fundamental ingredient
of wormhole geometries is the null energy condition violation,
phantom energy can provide  a means to support traversable
wormhole geometries \cite{Sushkov, Lobo2005, Zaslavskii2005}.
Indeed, due to the acceleration of  the universe, it seems
possible that macroscopic wormholes naturally  grow from
submicroscopic states that originally pervaded the quantum foam.
Moreover, it could be imagined an absurdly advanced civilization
mining the cosmic fluid for phantom energy necessary to construct
and sustain a traversable wormhole \cite{Lobo2005,
Zaslavskii2005}. Another point is that as the phantom energy
equation of state represents a spatially homogeneous cosmic fluid
and is assumed not to cluster, it is also possible that
inhomogeneities may arise due to gravitational instabilities.
Thus,  density fluctuations in the cosmological background may be
the origin of phantom wormholes. It can also be considered that
these structures are sustained by their own quantum fluctuations
\cite{Garattini1, Garattini2, Garattini3}.

Many of the papers published on the phantom energy
wormholes are not asymptotically flat \cite{Sushkov, Lobo2005,
Zaslavskii2005}. The approach of these papers is to glue the
interior wormhole metric to a vacuum exterior spacetime at a
junction interface \cite{Lemos1, Lobocqg, Crawford, Lemos2,
Lemos3}. Recently, new asymptotically flat phantom wormhole
solutions with no need to surgically pasting the interior wormhole
geometry to exterior vacuum spacetime have been found in
\cite{Lobo-Riazi}. On the other hand, spherically symmetric and static
traversable Morris-Thorne wormholes in the presence of a generic
cosmological constant $\Lambda$ are analyzed in \cite{Lemos4}. In
that paper, two spacetimes are glued into each other and explored
under matching conditions for the interior and exterior
spacetimes. Another paper in this direction is \cite{Rahaman} in
which the cosmological constant is considered as a space variable
scalar ($\Lambda = \Lambda(r)$).

In order to study the effects of the cosmological constant
background on the asymptotically flat wormholes, we use a special class of
wormholes solutions introduced by Lobo et al. \cite{Lobo-Riazi} which does
not need the cut and paste procedure.
We embed these asymptotically flat wormholes into
a generic cosmological constant background rather than a vacuum spacetime. Indeed,
there are  two 
inner and outer spacetimes like as the ones in the cut and paste procedure. But in this work and in contrast
to the cut and paste procedure, an inner asymptotically vanishing geometry (asymptotically
flat wormhole) is smoothly switching to the outer de Sitter or anti-de Sitter spacetime without need to a surgically pasting.  Near the throat, the wormhole
geometry is dominant while as the radii increases the wormhole features disappear
and the characteristic of the cosmological constant will be
more clarified, signalling us that our solutions include an
asymptotic spacetime, the background, which should normally be de
Sitter or anti-de Sitter, depending on the positive or negative
values of the cosmological constant. It is shown that for having
such a geometry, the wormhole throat $r_{0}$, cosmological
constant $\Lambda$ and the equation of state parameter $\omega$
must satisfy certain conditions. The organization of this paper is
as follows: In section II, general geometries and constraints of
Lorentzian wormholes are outlined. In section III, Einstein field
equations and the metric functions are studied and the above
mentioned conditions on $r_0$, $\Lambda$ and $\omega$ are
obtained. In sections IV and V, some specific solutions with their
mass function and energy-momentum tensor profiles are presented.
At the end of section V, the volume integral quantifier, for
general solutions obtained in section III, is briefly mentioned.
Finally, in section VI, we present our concluding remarks.
Throughout  this work, units of $G=c=1$ are used.
\section{General geometry and constraints of Lorentzian wormholes}
 The general static and spherically
symmetric Lorentzian wormhole metric is given by
\begin{equation}\label{21}
ds^{2}=-U(r)dt^{2}+\frac{dr^{2}}{1-\frac{b\left(r\right)}{r}}+r^{2}d\Omega^{2},
\end{equation}
where $d\Omega^{2}=d\theta^{2}+sin^{2}\theta d\phi^{2}$. The
metric functions $U(r)$ and $b(r)$ are referred to as the redshift
function and shape function, respectively. The general constraints
on the redshift and shape functions which build up a wormhole are
as follow:
\begin{enumerate}
\item The wormhole throat, which connects two asymptotic regions,
is located at the minimum radial coordinate $r_{0}$ at which
$b(r_{0})=r_{0}$. \item The shape function $b(r)$ must satisfy the
so-called {\it flaring-out condition}, which is valid at or near the throat vicinity, given by
\begin{equation}\label{22}
\frac{b(r)-rb^{\prime}(r)}{2b^{2}(r)}>0,
\end{equation}
which at the throat of the wormhole reduces to
$b^{\prime}(r_{0})<1$.
\item In order to keep the proper signature of
the metric, for the radial coordinates $r>r_{0}$, the shape
function should satisfy the condition
\begin{equation}\label{23}
1-\frac{b(r)}{r}>0.
\end{equation}

\item In order to have  asymptotically flat  geometries, the metric functions need
to obey the following conditions at $r\rightarrow\infty$ :
\begin{eqnarray}\label{24}
&&U(r)\rightarrow 1,\nonumber\\
&&\frac{b(r)}{r}\rightarrow 0.
\end{eqnarray}
Obviously, these conditions may be relaxed for no-asymptotically
flat wormholes. \item To ensure the absence of horizons and
singularities, it is also required that $U(r)$ be finite and
nonzero throughout the spacetime.

Notice that the above  constraints  provide a minimum set of
conditions which is mandatory for characterizing the  geometry of
two asymptotically flat regions connected by a bridge
\cite{Dadhich}. \textcolor{blue}{}
\end{enumerate}

\section{Einstein field equations and the metric functions}
 We consider an anisotropic fluid for the matter  content of the
 spacetime in the form of $T^{\mu}_{\nu}=diag(-\rho,p_{r},p_{l},p_{l})$
 where $\rho(r)$ represents the energy density, $p_{r}(r)$ is the radial
 pressure and $p_{l}(r)$ stands for the lateral pressure measured in the
 orthogonal direction to the radial direction. The Einstein equation with
 the cosmological constant $\Lambda$
\begin{equation}\label{35}
G_{\mu \nu}-\Lambda g_{\mu\nu}=8\pi T_{\mu\nu},
\end{equation}
leads to the following equations
\begin{equation}\label{36}
b^{\prime}(r)=\left(8\pi \rho(r) -\Lambda\right)r^{2},
\end{equation}
\begin{equation}\label{37}
\frac{U^{\prime} (r)}{U(r)}=\frac{8\pi p_{r}(r)r^{3}+\Lambda r^{3} +b(r)}{r\left(r-b(r)\right)},
\end{equation}
\begin{equation}\label{38}
p_{l}(r)=p_{r}(r)+\frac{r}{2}\left[p^{\prime}_{r}(r)+\left(\rho(r)
+p_{r}(r)\right)\frac{U^{\prime}(r)}{2U(r)}\right]
\end{equation}
where the prime sign denotes the derivative with respect to the
radial coordinate $r$. From equation (\ref{36}), it is apparent
that the total density $\rho(r)$ is as $\rho(r)=\rho_w
(r)+\rho_{\Lambda}$ where $\rho_w (r)=\frac{b^{\prime}(r)}{8 \pi
r^2}$ is the density profile induced by the wormhole structure and
$\rho_{\Lambda}\equiv\frac{\Lambda}{8\pi}$ is the density of the
cosmological constant $\Lambda$ which can be either positive or
negative representing the  de Sitter and anti-de Sitter regimes,
respectively.

One can define a  mass function $m(r)$  according to
\begin{equation}\label{39}
m(r)\equiv \int^{r}_{r_{0}}4\pi r^{2}\rho(r) dr,
\end{equation}

This equation together with the equation (\ref{36}) leads to
\begin{equation}\label{310}
m(r)=\frac{1}{2}\left[b(r)-r_{0}+\frac{\Lambda}{3}(r^{3}-r_{0}^{3})\right],
\end{equation}
which clearly vanishes at the wormhole throat $r=r_0$. Indeed, the
inclusion of the cosmological constant will shift the respective
values of $\rho(r)$, $p_{r}(r)$ and $p_{l}(r)$ and might help in
minimizing the amount of energy condition violating matter which
can be seen directly from equation (\ref{310}). Although the
corresponding mass of the cosmological constant is unbounded, the
wormhole part may or may not be bounded. We will consider these
two possibilities in coming next sections.

In this paper, we are interested in the wormhole solutions using
the barotropic equation of state $p_{r}(r)=\omega \rho(r)$. Thus,
using equations (\ref{36}) and (\ref{37}) we obtain
\begin{equation}\label{311}
\frac{U^{\prime}(r)}{U(r)}=\frac{r\omega b^{\prime}(r)+b(r)+\Lambda(1+\omega)r^{3}}{r\left(r-b(r)\right)}.
\end{equation}
Before going further, it is important to point out a subtlety in
considering the phantom energy equation of state in inhomogeneous
spherically symmetric wormhole spacetimes. As emphasized in
\cite{Sushkov} and \cite{Lobo2005, Zaslavskii2005}, the phantom
dark energy is a homogeneously distributed fluid with an isotropic
pressure. However, it can be extended to  the context of
inhomogeneous spacetimes by considering a negative radial pressure
in the equation of state. Then, the lateral pressure can be
obtained using equation (\ref{38}) which is coming from the
Einstein field equations. This approach is motivated by the
discussion of the inhomogeneities that may appear because of
gravitational instabilities and through the analysis carried out
in \cite{Sushkovb}. The authors of \cite{Sushkovb} investigated a
spherically symmetric time dependent wormhole solution in a
cosmological context with a ghost scalar field. As a result, the
radial pressure is negative through the spacetime and for large
values of the radial coordinate equals to the lateral pressure,
which shows the behavior of ghost scalar field as dark energy.

We have now four equations, equations (\ref{36})-(\ref{38}) and
(\ref{311}), with five unknown quantities  $U(r)$, $b(r)$,
$\rho(r)$, $p_{r}(r)$ and $p_{l}(r)$. There are two different
approaches for solving the field equations. One approach is to
consider a specific distribution of the energy density threading
the wormhole, like the approach of \cite{Sushkov} and consequently
finding the metric functions $U(r)$ and $b(r)$. The second
approach involves proposing a model wormhole geometry by imposing
specific choices for the shape and redshift functions and
obtaining the supporting energy-momentum tensor profile
\cite{Lobo2005, Zaslavskii2005}. In this paper, since we are
preliminary interested in finding asymptotically de Sitter and
anti- de Sitter wormhole solutions, which are supported by phantom
or non-phantom matter contents, the second approach is followed.

We consider wormholes with the shape function
\begin{equation}\label{312}
\frac{b(r)}{r_{0}}=a\left(\frac{r}{r_{0}}\right)^{\alpha}+C,
\end{equation}
where $a$, $\alpha$ and $C$ are dimensionless constants. The first
constraint imposes that $C=1-a$. In order to satisfy the fourth
constraint, we  get $\alpha<1$. Thus, the shape function takes the
form
\begin{equation}\label{313}
b(r)=r_{0}+ar_{0}\left[\left(\frac{r}{r_{0}}\right)^{\alpha}-1\right].
\end{equation}
In order to satisfy the flaring out condition we obtain
\begin{equation}\label{314}
1-a +a\left(\frac{r}{r_0}\right)^{\alpha}\left(1-\alpha\right)>0.
\end{equation}
Form the previously obtained result, $\alpha<1$, we may divide the
solutions of this condition into the following cases: i) $0\leq
a\leq 1$ which clearly satisfies the above condition,  ii) $a>1$
whose exact value depends on $\alpha$ and $r$ and iii) $a<0$ which
similar to the previous case, its exact value depending on
$\alpha$ and $r$. In addition, these three classes should also
satisfy the condition $a\alpha< 1$ coming from the flaring out
condition at the throat.

Using equations (\ref{36}) and (\ref{313}), we
can obtain the total energy density $\rho(r)$ as
\begin{equation}\label{315}
\rho(r)=\frac{1}{8\pi}\left(\frac{\alpha
a}{r_{0}^{2}}\left(\frac{r}{r_{0}}\right)^{\alpha -3}+\Lambda\right),
\end{equation}
where for the case of $\Lambda=0$, the profile density of
\cite{Lobo-Riazi} are recovered. Also, from equation (\ref{315}),
it is seen that in order to obtain de Sitter or anti-de Sitter
solutions when $r\rightarrow\infty$ we should have $\alpha<3$.
Then, our obtained restricted regime $\alpha<1$ includes these
asymptotic behaviors.

Also, the total energy density $\rho(r)$, equation (\ref{315}),
should satisfy the positive energy condition
\begin{equation}\label{316}
\frac{\alpha a}{r_{0}^{2}}\left(\frac{r}{r_{0}}\right)^{\alpha-3}+\Lambda\geq0,
\end{equation}
which is valid for $\Lambda>0$ with $\alpha a\geq0$. Then, with
respect to the above three classes of $a$ values and the condition
$a\alpha< 1$ coming from the flaring out condition at the throat,
we will have $0\leq\alpha a<1$ and the following classes are
distinguished:  i) $\Lambda>0$ with ranges of $0\leq a\leq 1$ and
$0\leq\alpha<1$, ii) $\Lambda>0$ with ranges of $1< a$ and
$0\leq\alpha<1$ and iii) $\Lambda>0$ with ranges of $a<0$ and
$\alpha\leq0$.

The condition (\ref{316}) can be satisfied for $\Lambda<0$ with
$0<\alpha a<1$ and  $\Lambda>0$ with $\alpha a<0$, but these cases
are completely  dependent on exact numerical  value of $\Lambda$.
The case of $\alpha a<0$  is arising from the presence of a
positive cosmological constant $\Lambda$ which is not allowed in
the absence of $\Lambda$ as in \cite{Lobo-Riazi}. Then, in the presence of
the cosmological constant, we have an extended class of the asymptotically
flat wormhole solutions than the ones in the absence of it. In addition,
when $\Lambda<0$, the positive energy condition for the energy
density $\rho(r)=\rho_w (r)+\rho_{\Lambda}\geq0$ can be violated.
The same situation occurs in the anti de-Sitter spacetime
\cite{Liddle}.

Returning to the equation (\ref{316}), we can obtain a condition
for the throat of the wormhole, $r=r_0$, when $\Lambda>0$ with the
range of $0\leq\alpha a<1$ as
\begin{equation}\label{317}
r_{0}^{2}\geq-\frac{\alpha a}{\Lambda},
\end{equation}
where it is trivial and does not put any restriction on the size
of the throat $r_0$. For the case of $\Lambda>0$ with $\alpha
a<0$, we also recover this condition but since $\alpha a$ has
negative values, it will be a nontrivial restriction on the size
of wormhole throat. If we consider $\Lambda<0$, we obtain
\begin{equation}\label{318}
r_{0}^{2}\leq-\frac{\alpha a}{\Lambda},
\end{equation}
where is a nontrivial restriction on throat size, since for this
case we should have just $0<\alpha a<1$. Equations (\ref{317}) and
(\ref{318}) reveal the dependence of the size of the wormhole
throat on the cosmological constant $\Lambda$ and shape function
parameters $a$ and $\alpha$.
In the absence of the cosmological constant, one may not deduce such a direct
result about the wormholes throat size in terms of its characterizing parameters
$a$ and $\alpha$ in equation (\ref{313}).

The condition of an event-horizon-free spacetime
requires that $U(r)$ to be finite and nonzero. Thus, due to the
finiteness of $U(r)$, as seen by
equation (7), the radial pressure evaluated at throat should be
  
\begin{equation}\label{319}
 p_{r}(r_{0})=-\frac{1}{8\pi}\left(\Lambda+\frac{1}{r_{0}^{2}}\right),
\end{equation}
Similar situation
arises also in the absence
of cosmological constant, as is shown  below equation (15) in reference [42]. Then, we have a negative pressure at the
throat which  provides the geometry with a repulsive character, preventing the wormhole
throat from collapsing. This characteristic is also exist for a positive cosmological
constant. For a negative cosmological constant, in order to have negative
pressure at the throat, we need $r_{0}^2<-\frac{1}{\Lambda}$ which can be consistent
with the obtained tighter constraint (18) coming from the positive energy condition.
In fact, the acceptable common range which satisfies both of these conditions, is as
equation (\ref{318}). Equation (\ref{319}) together with the equation of state  $p_{r}(r)=\omega
\rho(r)$ leads to
\begin{equation}\label{320}
\rho (r_{0})=-\frac{1}{8\pi \omega}\left(\Lambda +\frac{1}{r_{0}^{2}}\right).
\end{equation}
 
On the other hand, by substituting $r=r_0$ in equation (\ref{312})
we obtain
\begin{equation}\label{321}
\rho(r_{0})=\frac{1}{8\pi}\left(\frac{\alpha a}{r_{0}^{2}}+\Lambda\right),
\end{equation}
where the consistency between the two above equations gives the
cosmological constant as
 \begin{equation}\label{322}
 \Lambda=-\frac{1+ \alpha a \omega}{r_{0}^{2}\left(1+\omega\right)}.
 \end{equation}
In the absence of $\Lambda$ as in \cite{Lobo-Riazi}, for the
considered shape function (\ref{313}), the supporting matter with
$\omega=-\frac{1}{\alpha a}$ lies just in the phantom area with no lower
bound.   Thus, in order to obtain
solutions for $\Lambda>0$ with $0\leq \alpha a<1$, we should have
$-\frac{1}{\alpha a}<\omega<-1$ which points to a restricted
phantom era with a lower bound specified by given $\alpha$ and $a$
values. Also, when we have $\Lambda<0$ and $0< \alpha a<1$, there
are accessible solutions for both  ranges $\omega<-\frac{1}{\alpha
a}<-1$ and $\omega>-1$ which correspond the phantom and
non-phantom regimes, respectively.
The solutions with  non-phantom regimes will be naturally ruled out by the
flaring out condition. It is well known that the flaring out condition, equation (\ref{22}), leads
to $\rho + p_r<0 $ for the fluid near a wormhole throat which demands the matter fields with $\omega<-1$, the phantom matters. This is the well-known violation of the null and weak energy conditions.
Equation (\ref{322}) gives a direct constraint for the throat
size of the wormholes living in a generic cosmological background. For 
given values of the cosmological constant, the shape function characterizing
parameters $a$ and $\alpha$ with equation of state parameter of the wormhole supporting matters $\omega$, the throat size will be fixed. For a wormhole
embedded in a vacuum spacetime, one can not deduce such a direct constraint
on its throat size.

Considering the shape function given by equation (\ref{313}), the
ordinary differential equation for the redshift function
(\ref{311}) takes the following form
\begin{equation}\label{323}
\frac{U^{\prime} (r)}{U(r)}=\frac{\Lambda\left(1+\omega\right)r}{1-\left(\frac{r_{0}}{r}\right)
\left(1+a\left[(\frac{r}{r_{0}})^{\alpha}-1 \right]\right)}+\left(\frac{r_{0}}{r^{2}}\right)\frac{1+a
\left[\left(\frac{r}{r_0}\right)^{\alpha}\left(1+\alpha \omega\right)-1\right]}{1-\left(\frac{r_{0}}{r}
\right)\left(1+a\left[\left(\frac{r}{r_{0}}\right)^{\alpha}-1 \right]\right)},
\end{equation}
where the results of \cite{Lobo-Riazi} are simply recovered by
substituting $\Lambda=0$. On the other hand, equations (\ref{313})
and (\ref{315}) can be used to write the shape function as
\begin{eqnarray}\label{324}
&b(r)&=\frac{8\pi}{\alpha}\rho(r)r^{3}-\frac{\Lambda}{\alpha}r^{3}+r_{0}\left(1-a\right)\nonumber\\
&&=\frac{8\pi}{\alpha \omega}p_{r}(r)r^{3}-\frac{\Lambda}{\alpha}r^{3}+r_{0}\left(1-a\right).
\end{eqnarray}
It is seen that this equation reveals the de Sitter or anti-de
Sitter background of the whole spacetime.  Note that equation (\ref{324}) does not contradict with the
asymptotically flatness of the wormhole geometry, because of the matter density
obtained in equation (\ref{315}). In fact, the considered
asymptotic flat form  for the shape function, equation
(\ref{313}), affects the amount of energy condition violating
matter and couples the wormhole throat size $r_0$, cosmological
constant $\Lambda$ and equation of state parameter $\omega$ to
each other as obtained in equations (\ref{317}), (\ref{318}) and
(\ref{322}).  

Consequently, one is able to substitute equation
(\ref{322}) into equation (\ref{311}) and after using equation
(\ref{36}), obtain the following
\begin{equation}\label{325}
\frac{U^{\prime} (r)}{U(r)}=\frac{8\pi p_{r}\left(\frac{1+\alpha\omega}{\alpha\omega}\right)r^{3}
+\Lambda\left(\frac{\alpha -1}{\alpha}\right)r^{3}+r_{0}(1-a)}{r^{2}\left(1-\frac{8\pi}{\alpha \omega}
p_{r}r^{2}+\frac{\Lambda}{\alpha}r^{2}-\frac{r_{0}}{r}\left(1-a\right)\right)}.
\end{equation}
Unfortunately, equations (\ref{323}) and (\ref{325}) in general
have not an exact solution. Thus, in order to deduce exact
wormhole solutions for this equations, we will consider some
specific choices for the parameters $a$ and $\alpha$ in the next
sections.

Meanwhile, using equations (\ref{324}) and (\ref{325}) we can
rewrite the lateral pressure (\ref{38}) in a general form as
\begin{eqnarray}\label{326}
&p_{l}(r)=&p_{r}(r)\left(\frac{\alpha-1}{2}\right)+\frac{\Lambda \omega\left(3-\alpha \right)}{16\pi}\nonumber\\
&&+p_{r}(r)\left(\frac{1-\alpha a }{4(\Lambda r_{0}^{2}+1)}\right)\frac{8\pi p_{r}(r)\left(\frac{\alpha\omega+1}
{\alpha\omega}\right)r^{2}+\Lambda\left(\frac{\alpha-1}{\alpha}\right)r^{2}+\frac{r_{0}}{r}\left(1-a\right)}{1-\frac{8\pi}{\alpha \omega}
p_{r}(r)r^{2}+\frac{\Lambda}{\alpha}r^{2}-\frac{r_{0}}{r}\left(1-a\right)},
\end{eqnarray}
where the first and second terms are due to the pure mass and pure
cosmological constant effects, respectively, while the third term
is a mixed term.

Obviously, in the presence of cosmological constant, the energy
density profile $\rho(r)$, equation (\ref{315}), have an
asymptotically de Sitter or anti de Sitter behavior as
$r\rightarrow \infty$. Thus, the radial pressure will take an
asymptotically de Sitter or anti de Sitter behavior $p_r
\rightarrow\frac{\omega\Lambda}{8\pi}$ as $r\rightarrow\infty$
where $\omega$ should be $-1$ at spatial infinity. For
the lateral pressure,  the second term in brackets in equation (\ref{38}) with respect to equations (\ref{315}) and (\ref{325}) will vanish at $r\rightarrow\infty$.
Then, the lateral pressure also will take an
asymptotically de Sitter or anti de Sitter behavior $p_l
\rightarrow\frac{\omega\Lambda}{8\pi}$ as $r\rightarrow\infty$
where $\omega$ should be $-1$ at spatial infinity. 

Also, one may consider a constant
redshift function for simplicity in which case the de Sitter or anti-de
Sitter asymptotics are simply achieved \cite{Cataldo}. Since any constant
redshift function can be absorbed in the re-scaled time
coordinate, we will consider $U(r)=1$. This consideration
along with Eq.~(\ref{313}), guarantees the asymptotically flatness
condition for the inner spacetime (the wormhole spacetime).
Finally, we recover the field equations as follows
\begin{equation}\label{327}
\frac{b^{\prime}(r)}{r^2}=8\pi \rho(r) -\Lambda,
\end{equation}
\begin{equation}\label{328}
-\frac{b}{r^3}=8\pi p_{r}(r)+\Lambda  ,
\end{equation}
\begin{equation}\label{329}
\frac{b-b^{\prime}r}{2r^3}=8\pi p_{t}(r)+\Lambda.
\end{equation}
Clearly, using the shape function $b(r)$, equation (\ref{313}), we
have de Sitter or anti-de Sitter asymptotics as $\rho\rightarrow
\frac{\Lambda}{8\pi}$, $p_r\rightarrow -\frac{\Lambda}{8\pi}$ and
$p_t\rightarrow -\frac{\Lambda}{8\pi}$ as $r\rightarrow\infty$.
Also, for the case of constant redshift function, one may go
further and consider equation of state $p_r =\omega\rho$ and using
equations (\ref{327}) and (\ref{328}) obtains the space varying
equation of state parameter $\omega$ as
\begin{equation}\label{330}
\omega(r)=-\frac{\Lambda r^3 +b }{\Lambda r^3 + rb^{\prime}},
\end{equation}
which  help us to achieve a better understanding of the behavior of
the dominant fluid and reveals de Sitter or anti-de Sitter nature
of spacetime as $\omega\rightarrow-1$ as $r\rightarrow\infty$.
This equation leads to  equation (\ref{322}) at the throat of the
wormhole and shows that $\omega$ as a space varying parameter is
not allowed to be $\omega=-1$ at the throat of the wormhole. The
generalization of equation (\ref{330}) to the case of $U(r)\neq
constant $ is also applicable by using equations (\ref{36}) and
(\ref{37}).

As we saw, while cosmological constant affects the
wormhole and its properties, the wormhole geometry will disappear
in the $r \rightarrow \infty$ limit. We should also note that the
asymptotically flatness condition for the inner spacetime leads to
$g_{\alpha \beta}\rightarrow \eta_{\alpha \beta}$ and $G_{\alpha
\beta}\rightarrow 0$ yielding to $-\Lambda \eta_{\alpha \beta}
\sim 8\pi T_{\alpha \beta}$, where $\eta_{\alpha
\beta}=diag(-1,1,r^2,r^2 \sin^2\theta)$ is the flat spacetime
metric. Thus, we see that the de Sitter and anti-de Sitter
spacetimes will be dominated in the $r \rightarrow \infty$ limit. Indeed,
there are  two 
inner and outer spacetimes like as the ones in the cut and paste procedure. But in this work and in contrast
to the cut and paste procedure, an inner asymptotically vanishing geometry (asymptotically
flat wormhole) is smoothly switching to the outer de Sitter or anti-de Sitter spacetime. Also, One can only consider one metric by imposing the
asymptotically  de Sitter or anti-de Sitter conditions on the
metric and use $G_{\alpha \beta}=8 \pi T_{\alpha \beta}$ as the
Einstein equation. This situation is like what we have in the
Schwarzschild-de Sitter spacetime \cite{sds}, which is also different
than our approach in this work.

Finally, using the shape function, equation (\ref{313}), we can
express the third wormhole constraint mentioned in section II, by
the following inequality
\begin{equation}\label{331}
 H(x,a,\alpha)\equiv
a x^{-\alpha}-x^{-1}+1-a<0,
\end{equation}
where we defined $x\equiv \frac{r_{0}}{r}$. In order to cover the entire
spacetime, the $x$ has the range of $0<x\leq 1$ where $x=1$
corresponds to the wormhole throat, $r=r_{0}$, and $x\rightarrow
0$ corresponds to spatial infinity. In order to check this condition,
we shall plot $H(x,a, \alpha)$ versus $x\equiv \frac{r_{0}}{r}$
for some specific choices of $a$ and $\alpha$ at  the end of the next section.
Also, with note to the importance of the amount of energy violating matter, we will classify our obtained exact solutions to the wormholes with
an unbounded or a bounded mass function. For the wormholes with a bounded mass function, a finite "amount of energy conditions violating matter" is sufficient in order to support the corresponding asymptotically flat
wormhole geometry. 
\section{Specific Wormholes with an unbounded mass function}
 \subsection{The case  $a=1$ and $\alpha=\frac{1}{2}$}
This case is allowed for both $\Lambda>0$ and $\Lambda<0$.
For this specific case, equation (\ref{323}) can be solved:
\begin{equation}\label{432}
U(r)=U_{1}exp\left[-\frac{\Lambda\left(3r^{2}+6r_{0}r+4r_{0}r\left(\frac{r}{r_0}\right)
^{\frac{1}{2}}+12r_{0}^{2}\left(\frac{r}{r_0}\right)^{\frac{1}{2}}+
6r_{0}^{2}\ln\left(\frac{r}{r_0}\right)\right)}{12\Lambda
r_{0}^{2}+6}\right],
\end{equation}
where we can absorb the constant $U_1$ into the re-scaled time coordinate.
The mass function takes the simple form
\begin{equation}\label{433}
m(r)=\frac{r_{0}}{2}\left[\left(\frac{r}{r_0}\right)^{\frac{1}{2}}-1\right]+\frac{\Lambda}{6}\left(r^{3}-r_{0}^{3}\right),
\end{equation}
and the pressures will be
\begin{eqnarray}\label{434}
&&p_r(r)=\omega\rho(r)=\frac{\omega}{8\pi}\left[\frac{1}{2r_{0}^{2}}\left(\frac{r_0}{r}\right)^{\frac{5}{2}}+\Lambda\right],\nonumber\\
&&p_{l}(r)=-\frac{1}{4}p_{r}+\frac{5\Lambda\omega}{32\pi}+\frac{ 8p_{r}\left(\frac{\omega+2}{\omega}\right)\pi
r^{2}-\Lambda
r^{2}}{8\left(\Lambda r_{0}^{2}+1\right)\left(1-\frac{16\pi}{\omega}p_{r}r^{2}+2\Lambda r^{2}\right)}p_{r}.
\end{eqnarray}

Also, by considering the constant redshift
function $U(r)=1$ ,  we recover $m(r)$ in equation (\ref{433}) and
the pressures will be
\begin{eqnarray}\label{435}
&&p_r(r)=-\frac{1}{8\pi}\left[\frac{r_0}{r^3}\sqrt{\frac{r}{r_0}}+\Lambda\right],\nonumber\\
&&p_{l}(r)=\frac{1}{8\pi}\left[\frac{1}{4r^2}\sqrt{\frac{r_0}{r}}-\Lambda\right],
\end{eqnarray}
which have the asymptotics as $r\rightarrow\infty$ as
\begin{eqnarray}\label{436}
&&p_r(r)\rightarrow-\frac{\Lambda}{8\pi},\nonumber\\
&&p_{l}(r)\rightarrow-\frac{\Lambda}{8\pi},
\end{eqnarray}
corresponding to the asymptotically de Sitter or anti-de Sitter spacetimes.

\subsection{The case  $a=1$ and $\alpha=\frac{1}{3}$}
This case is also allowed for both $\Lambda>0$ and $\Lambda<0$. For the specific case $a=1$ and $\alpha=\frac{1}{3}$, solving the
equation (\ref{323}) gives the solution
\begin{equation}\label{437}
U(r)=U_{2}exp\left[-\frac{\Lambda\left(2r^{2}+3r_{0}r\left(\frac{r}{r_0}\right)^{\frac{1}{3}}+6r_{0}^{2}
\left(\frac{r}{r_0}\right)^{\frac{2}{3}}+4r_{0}^{2}\ln\left(\frac{r}{r_0}\right)\right)}{6\Lambda
r_{0}^{2}+2}\right],
\end{equation}
where we can treat $U_2$
as previous section.

Also, for the mass function, we will have
\begin{equation}\label{438}
m(r)=\frac{r_{0}}{2}\left[\left(\frac{r}{r_{0}}\right)^{\frac{1}{3}}-1\right]+\frac{\Lambda}{6}\left(r^{3}-r_{0}^{3}\right).
\end{equation}

The pressures also will be
\begin{eqnarray}\label{439}
&&p_r(r)=\omega\rho(r)=\frac{\omega}{8\pi}\left[\frac{1}{3r_{0}^{2}}\left(\frac{r_0}{r}\right)^{\frac{8}{3}}+\Lambda\right],\nonumber\\
&&p_{l}(r)=-\frac{1}{3}p_{r}+\frac{\Lambda\omega}{6\pi}+\frac{4\pi p_{r}\left(\frac{\omega+3}{\omega}\right)r^{2}-\Lambda
r^{2}}{3\left(\Lambda r_{0}^{2}-1\right)\left(1-\frac{24\pi}{\omega}p_{r}r^{2}+3\Lambda r^{2}\right)}p_{r}.
\end{eqnarray}
Also, by considering the constant redshift
function $U(r)=1$ ,  we recover $m(r)$ in equation (\ref{433}) and
the pressures will be
\begin{eqnarray}\label{440}
&&p_r(r)=-\frac{1}{8\pi}\left[\frac{r_0}{r^3}\left(\frac{r}{r_0}\right)^{\frac{1}{3}}+\Lambda\right],\nonumber\\
&&p_{l}(r)=\frac{1}{8\pi}\left[\frac{1}{3r^2}\left(\frac{r_0}{r}\right)^{\frac{2}{3}}-\Lambda\right],
\end{eqnarray}
which have the asymptotics as $r\rightarrow\infty$ as
\begin{eqnarray}\label{441}
&&p_r(r)\rightarrow-\frac{\Lambda}{8\pi},\nonumber\\
&&p_{l}(r)\rightarrow-\frac{\Lambda}{8\pi},
\end{eqnarray}
 corresponding to the asymptotically de Sitter or anti-de Sitter spacetimes.

\subsection{The case  $a=\frac{1}{2}$ and $\alpha=\frac{1}{2}$}
This case is also acceptable for  $\Lambda>0$ and $\Lambda<0$. For this configuration, one obtains the following solution
\begin{eqnarray}\label{442}
&U(r)=&U_{3}exp\left[-\frac{\Lambda\left(12r^{2}+18r_{0}r+8r_{0}r\left(\frac{r}{r_0}\right)^{\frac{1}{2}}
+30r_{0}^{2}\left(\frac{r}{r_0}\right)^{\frac{1}{2}}+32r_{0}^{2}\ln\left(\frac{r}{r_0}\right)-31r_{0}^{2}
\ln\left(2\left(\frac{r}{r_0}\right)^{\frac{1}{2}}+1\right)\right)}{32\Lambda
r_{0}^{2}+8}\right]\nonumber\\
&&\times \left[ 4-4\sqrt \frac{r_0}{r}- \frac{r_0}{r}\right]^{\frac{1}{4\Lambda
r_{0}^2}},
\end{eqnarray}
where again, absorption of the constant $U_3$ would be applicable
into the re-scaled time coordinate.

The mass function will be written as
\begin{equation}\label{443}
m(r)=\frac{r_{0}}{4}\left[\left(\frac{r}{r_{0}}\right)^{\frac{1}{2}}-1\right]+\frac{\Lambda}{6}\left(r^{3}-r_{0}^{3}\right),
\end{equation}
while the pressures are obtained as
\begin{eqnarray}\label{444}
&&p_r(r)=\omega\rho(r)=\frac{\omega}{8\pi}\left[\frac{1}{4r_{0}^{2}}\left(\frac{r_0}{r}\right)^{\frac{5}{2}}+\Lambda\right],\nonumber\\
&&p_{l}(r)=-\frac{1}{4}p_{r}+\frac{5\Lambda\omega}{32\pi}+\frac{24\pi p_{r}\left(\frac{\omega+2}{\omega}\right)r^{2}-3\Lambda
r^{2}+\frac{3r_0}{2r}}{16\left(\Lambda r_{0}^{2}+1\right)\left(1-\frac{16\pi}{\omega}p_{r}r^{2}+2\Lambda r^{2}-\frac{r_0}{2r}\right)}p_{r}.
\end{eqnarray}
Also, by considering the constant redshift
function $U(r)=1$ ,  we recover $m(r)$ in equation (\ref{433}) and
the pressures will be
\begin{eqnarray}\label{445}
&&p_r(r)=-\frac{1}{8\pi}\left[\frac{r_0}{2r^3}\left(1+(\frac{r}{r_0})^{\frac{1}{2}}\right)+\Lambda\right],\nonumber\\
&&p_{l}(r)=\frac{1}{8\pi}\left[\frac{r_{0}}{8r^3}\left(2+(\frac{r}{r_0})^{\frac{1}{2}}\right)-\Lambda\right],
\end{eqnarray}
which have the asymptotics as $r\rightarrow\infty$ as
\begin{eqnarray}\label{446}
&&p_r(r)\rightarrow-\frac{\Lambda}{8\pi},\nonumber\\
&&p_{l}(r)\rightarrow-\frac{\Lambda}{8\pi},
\end{eqnarray}
corresponding to the asymptotically de Sitter or anti-de Sitter spacetimes.
\subsection{The case  $a=-1$ and $\alpha=\frac{1}{2}$}
This case is only allowed for   $\Lambda>0$. Applying the values $a=-1$ and $\alpha=\frac{1}{2}$, gives a
solution for the equation (\ref{323}) as
\begin{eqnarray}\label{447}
&U(r)=U_{5}\times\nonumber\\
&exp\left[-\frac{\Lambda\left(3r^{2}-60r_{0}^{2}\left(\frac{r}{r_0}\right)^{\frac{1}{2}}+
18r_{0}r+124r_{0}^{2}\ln\left(\left(\frac{r}{r_0}\right)^{\frac{1}{2}}+2\right)+
4r_{0}^{2}\ln\left(\frac{r}{r_0}\right)
-4r_{0}r\left(\frac{r}{r_0}\right)^{\frac{1}{2}}\right)-2\ln\left(\frac{r}{r_0}\right)
+4\ln\left(\left(\frac{r}{r_0}\right)^{\frac{1}{2}}+2)\right)}
{4\Lambda r_{0}^{2}-2}\right],
\end{eqnarray}
where the constant $U_4$ can be absorbed similar to the previous solutions.

The mass function would be written as
\begin{equation}\label{448}
m(r)=\frac{r_{0}}{2}\left[1-\left(\frac{r}{r_0}\right)^{\frac{1}{2}}\right]+
\frac{\Lambda}{6}\left(r^{3}-r_{0}^{3}\right),
\end{equation}
and the pressures  will be
\begin{eqnarray}\label{449}
&&p_r(r)=\omega\rho(r)=\frac{\omega}{8\pi}\left[-\frac{1}{2r_{0}^{2}}\left(\frac{r_0}{r}\right)^{\frac{5}{2}}+\Lambda\right],\nonumber\\
&&p_{l}(r)=-\frac{1}{4}p_{r}+\frac{5\Lambda\omega}{32\pi}+\frac{24\pi p_{r}\left(\frac{\omega+2}{\omega}\right)r^{2}-3\Lambda
r^{2}+6\frac{r_0}{r}}{8\left(\Lambda r_{0}^{2}+1\right)\left(1+\frac{16\pi}{\omega}p_{r}r^{2}+2\Lambda r^{2}-2\frac{r_0}{r}\right)}p_{r}.
\end{eqnarray}
Also, by considering the constant redshift
function $U(r)=1$ ,  we recover $m(r)$ in equation (\ref{433}) and
the pressures will be
\begin{eqnarray}\label{450}
&&p_r(r)=-\frac{1}{8\pi}\left[\frac{r_0}{r^3}\left(2-(\frac{r}{r_0})^{\frac{1}{2}}\right)+\Lambda\right],\nonumber\\
&&p_{l}(r)=\frac{1}{8\pi}\left[\frac{r_{0}}{4r^3}\left(4-(\frac{r}{r_0})^{\frac{1}{2}}\right)-\Lambda\right],
\end{eqnarray}
which have the asymptotics as $r\rightarrow\infty$ as
\begin{eqnarray}\label{451}
&&p_r(r)\rightarrow-\frac{\Lambda}{8\pi},\nonumber\\
&&p_{l}(r)\rightarrow-\frac{\Lambda}{8\pi},
\end{eqnarray}
corresponding to the asymptotically de Sitter or anti-de Sitter
spacetimes.

\section{Specific wormholes with a bounded mass function}
\subsection{The case $a=-\frac{1}{2}$ and $\alpha=-1$}
This case is allowed for both  $\Lambda>0$ and $\Lambda<0$.
Applying the values $a=-\frac{1}{2}$ and $\alpha=-1$, gives a
solution for equation (\ref{323}) as
\begin{equation}\label{452}
U(r)=U_{4}exp\left[-\frac{\Lambda\left(2r^{2}
+6r_{0}r-9r_{0}^{2}\ln\left(2r-r_0\right)+16r_{0}^{2}\ln\left(r\right)\right)-
12\ln\left(2r-r_0\right)+12\ln\left(r\right)}{8\Lambda
r_{0}^{2}+4}\right],
\end{equation}
where the constant $U_4$ can be absorbed similar to the previous solutions.

The mass function would be written as
\begin{equation}\label{453}
m(r)=\frac{r_{0}}{4}\left[1-\frac{r_0}{r}\right]+\frac{\Lambda}{6}\left(r^{3}-r_{0}^{3}\right).
\end{equation}

The pressures will be
\begin{eqnarray}\label{454}
&&p_r(r)=\omega\rho(r)=\frac{\omega}{8\pi}\left[\frac{1}{2r_{0}^{2}}\left(\frac{r_0}{r}\right)^{4}
+\Lambda\right],\nonumber\\
&&p_{l}(r)=-p_{r}+\frac{\Lambda\omega}{4\pi}+\frac{8\pi p_{r}\left(\frac{\omega-1}{\omega}\right)r^{2}+2\Lambda
r^{2}+\frac{3r_0}{2r}}{8\left(\Lambda r_{0}^{2}+1\right)\left(1+\frac{8\pi}{\omega}p_{r}r^{2}-\Lambda r^{2}-\frac{3r_0}{2r}\right)}p_{r}.
\end{eqnarray}
Also, by considering the constant redshift
function $U(r)=1$ ,  we recover $m(r)$ in equation (\ref{433}) and
the pressures will be
\begin{eqnarray}\label{455}
&&p_r(r)=-\frac{1}{8\pi}\left[\frac{r_0}{2r^4}(3r-r_0)+\Lambda\right],\nonumber\\
&&p_{l}(r)=\frac{1}{8\pi}\left[\frac{r_{0}}{4r^4}\left(3r-2r_0\right)-\Lambda\right],
\end{eqnarray}
which have the asymptotics as $r\rightarrow\infty$ as
\begin{eqnarray}\label{456}
&&p_r(r)\rightarrow-\frac{\Lambda}{8\pi},\nonumber\\
&&p_{l}(r)\rightarrow-\frac{\Lambda}{8\pi},
\end{eqnarray}
corresponding to the asymptotically de Sitter or anti-de Sitter
spacetimes.

\subsection{The case  $a=\frac{1}{2}$ and $\alpha=-1$}
This case is only allowed for  $\Lambda>0$. For the specific case $a=\frac{1}{2}$ and $\alpha=-1$, solving the
equation (\ref{323}) gives the solution
\begin{equation}\label{457}
U(r)=U_{2}exp\left[-\frac{\Lambda\left(6r^{2}+6r_{0}r
-7r_{0}^{2}\ln\left(2r+r_{0}\right)
+16r_{0}^{2}\ln\left(r\right)\right)-4\ln\left(2r+r_{0}\right)+4\ln\left(r\right)}{8\Lambda
r_{0}^{2}-4}\right],
\end{equation}
where we can treat $U_2$
as previous sections.

For the mass function, we have
\begin{equation}\label{458}
m(r)=\frac{r_{0}}{4}\left[\left(\frac{r_0}{r}\right)-1\right]+
\frac{\Lambda}{6}\left(r^{3}-r_{0}^{3}\right).
\end{equation}

The pressures will be
\begin{eqnarray}\label{459}
&&p_r(r)=\omega\rho(r)=\frac{\omega}{8\pi}\left[-\frac{1}{2r_{0}^{2}}\left(\frac{r_0}{r}\right)^{4}+\Lambda\right],\nonumber\\
&&p_{l}(r)=-p_{r}+\frac{\Lambda\omega}{4\pi}+\frac{24\pi p_{r}\left(\frac{\omega-1}{\omega}\right)r^{2}+6\Lambda
r^{2}+\frac{3r_0}{2r}}{8\left(\Lambda r_{0}^{2}+1\right)\left(1+\frac{8\pi}{\omega}p_{r}r^{2}-\Lambda r^{2}-\frac{r_0}{2r}\right)}p_{r}.
\end{eqnarray}
Also, by considering the constant redshift
function $U(r)=1$ ,  we recover $m(r)$ in equation (\ref{433}) and
the pressures will be
\begin{eqnarray}\label{460}
&&p_{r}(r)=-\frac{1}{8\pi}\left[\frac{r_0}{2r^4}(r+r_0)+\Lambda\right],\nonumber\\
&&p_{l}(r)=\frac{1}{8\pi}\left[\frac{r_{0}}{4r^4}\left(r+2r_{0}\right)-\Lambda\right],
\end{eqnarray}
which have the asymptotics as $r\rightarrow\infty$ as
\begin{eqnarray}\label{461}
&&p_r(r)\rightarrow-\frac{\Lambda}{8\pi},\nonumber\\
&&p_{l}(r)\rightarrow-\frac{\Lambda}{8\pi},
\end{eqnarray}
corresponding to the asymptotically de Sitter or anti-de Sitter
spacetimes.

The corresponding $H(x,a,\alpha)$ function for all of these cases
are shown in the following figure, Figure 1.
\begin{figure}[ht]
\centering
\includegraphics[scale=0.4]{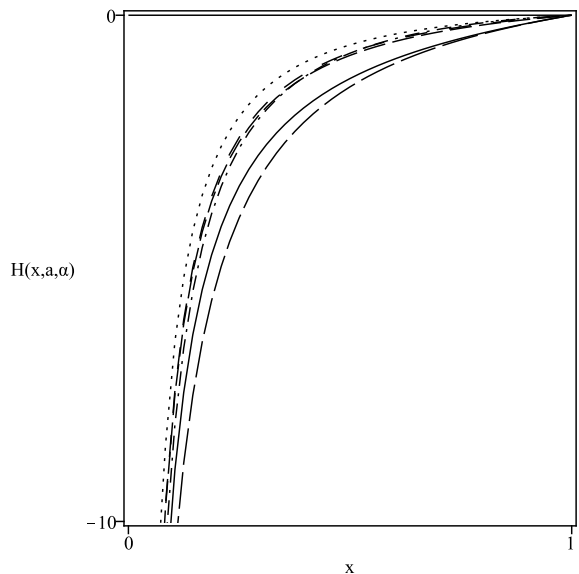}
\caption{The plots depict $H(x, a, \alpha)$ function in which the
solid, dot, spacedash, dash, longdash and dashdot plots stand for  the $H(x, 1, 1/2)$,
$H(x, 1, 1/3)$, $H(x, 1/2, 1/2)$, $H(x, -1, 1/2)$, $H(x, -1/2, -1)$ and $H(x, 1/2, -1)$,
respectively. The parameter $x = r_{0}/r$, lying in the range
$0<x\leq 1$,
has been defined in order to cover the entire spacetime.}
\end{figure}

As seen from the figure, the function $H(x, a, \alpha)$ is
negative for all cases throughout the entire range of $x$,
indicating the satisfaction of the third wormhole condition.

Also, it is interesting to evaluate the " volume integral
quantifier " \cite{Visser, Kar} which provides information about
the " total amount " of energy condition violating matter. This
quantity is given by
\begin{equation}\label{463}
I_{V}\equiv  \oint [\rho(r)+p_{r}(r)]dV
=  2\int^{\infty}_{r_0} [\rho(r)+p_{r}(r)]4\pi r^{2}dr,
\end{equation}
which by considering the equation (\ref{318}) and the equation of
state $p_{r}(r)=\omega\rho(r)$ gives the solution as
\begin{equation}\label{464}
I_{V}=(\omega +1)\left[ar_{0}(\frac{r}{ r_0})^{\alpha}+\frac{1}{3}\Lambda r^{3}\right]\mid_{r_0}^{\infty}.
\end{equation}
This equation shows that in the absence of cosmological constant,
in order to have a finite amount of "energy condition violating
matter", the $\alpha$ values must be negative, in which case we
have
\begin{equation}
I_{V}\rightarrow -(\omega + 1)ar_{0}.
\end{equation}
Therefore, as $a\rightarrow 0$, we have $I_{V}\rightarrow 0$ which
reflects arbitrary small quantities of energy condition violating
matter. In the presence of cosmological constant, it is seen that
the sign of equation (\ref{450}) is not fixed and depends on the
parameters of the model and cosmological constant.

\section{CoNCLUDING Remarks}
The asymptotically flat wormhole
solutions embedded in a generic cosmological constant background rather than a vacuum
spacetime are investigated. Indeed,
there are  two 
inner and outer spacetimes like as the ones in the cut and paste procedure. But in this work and in contrast
to the cut and paste procedure, an inner asymptotically vanishing geometry (asymptotically
flat wormhole) is smoothly switching to the outer de Sitter or anti-de Sitter spacetime without need to a surgically pasting.  Near the throat, the wormhole
geometry is dominant while as the radii increases the wormhole features disappear
and the characterizations of the cosmological constant will be
more clarified, signalling us that our solutions include an
asymptotic spacetime, the background, which should normally be de
Sitter or anti-de Sitter, depending on the positive or negative
values of the cosmological constant. It is shown that for constructing such a geometry, the wormhole
throat $r_{0}$, cosmological constant $\Lambda$ and the equation
of state parameter $\omega$ should satisfy certain relations. With
respect to the sign of $\Lambda$, corresponding to de Sitter or
anti-de Sitter spacetime, a new restricting condition on wormhole
throat size is obtained. Also, it is shown that in the presence
of cosmological constant $\Lambda$ with a general redshift
function $U(r)$, the energy density profile $\rho(r)$, the
radial pressure $p_{r}(r)$ and the lateral pressure $p_l (r)$ take an asymptotically de Sitter or
anti de Sitter behavior  at $r\rightarrow\infty$. Also, it is denoted that
these asymptotics are simply achieved by  choosing  a constant
redshift function.  Then, using the possibility of
setting different values for the parameters of the model, 
some exact solutions leading to specific metrics, mass functions
and supporting energy momentum profiles are found. The volume integral quantifier, which provides useful information
about the total amount of energy condition violating matter is also briefly
mentioned. It
is shown that the amount of this energy condition violation
depends on the parameters of the model and the value of
cosmological constant which can be fixed from observations.

\textbf{}

\section*{Acknowledgment}
Y. Heydarzade acknowledges F. Parsaei and R. Monadi for checking some
calculations and useful comments.


\end{document}